# Gaussian Statistics of the Cosmic Microwave Background: Correlation of Temperature Extrema In the *COBE* [1] DMR Two-Year Sky Maps

A. Kogut[2,3], A.J. Banday[4], C.L. Bennett[5], G. Hinshaw[2],
P.M. Lubin[6], and G.F. Smoot[7]

## ABSTRACT

We use the two-point correlation function of the extrema points (peaks and valleys) in the *COBE* Differential Microwave Radiometers (DMR) 2-year sky maps as a test for non-Gaussian temperature distribution in the cosmic microwave background anisotropy. A maximum likelihood analysis compares the DMR data to $n = 1$ toy models whose random-phase spherical harmonic components $a_{\ell m}$ are drawn from either Gaussian, $\chi^2$, or log-normal parent populations. The likelihood of the 53 GHz (A+B)/2 data is greatest for the exact Gaussian model. All non-Gaussian models tested are ruled out at 90% confidence, limited by type II errors in the statistical inference. The extrema correlation function is a stronger test for this class of non-Gaussian models than topological statistics such as the genus.

*Subject headings:* cosmic microwave background — methods:statistical

---

[1] The National Aeronautics and Space Administration/Goddard Space Flight Center (NASA/GSFC) is responsible for the design, development, and operation of the Cosmic Background Explorer (*COBE*). Scientific guidance is provided by the *COBE* Science Working Group. GSFC is also responsible for the development of analysis software and for the production of the mission data sets.

[2] Hughes STX Corporation, Laboratory for Astronomy and Solar Physics, Code 685, NASA/GSFC, Greenbelt MD 20771

[3] E-mail: kogut@stars.gsfc.nasa.gov

[4] Universities Space Research Association, Laboratory for Astronomy and Solar Physics, Code 685.9, NASA/GSFC, Greenbelt MD 20771

[5] Laboratory for Astronomy and Solar Physics, NASA Goddard Space Flight Center, Code 685, Greenbelt MD 20771

[6] UCSB Physics Department, Santa Barbara CA 93106

[7] LBL, SSL, & CfPA, Bldg 50-351, University of California, Berkeley CA 94720

# 1. Introduction

The angular distribution of the cosmic microwave background (CMB) probes the distribution of mass and energy in the early universe and provides a means to test competing models of structure formation. One such test is whether or not the distribution of CMB anisotropies follows Gaussian statistics. In most inflationary models, the large-scale CMB anisotropy results from quantum fluctuations and follow their Gaussian statistics. Competing models (topological defects, axions, late phase transitions) generally involve higher-order correlations and produce non-Gaussian distributions. Attempts to differentiate Gaussian from non-Gaussian distributions on large angular scales are complicated by the tendency of any distribution to approach Gaussian when averaged over a sufficiently large area (the central limit theorem) and by our inability to measure more than one sample (our observable universe) of the theoretical parent distribution ("cosmic variance").

Several authors (Hinshaw et al. 1994, Smoot et al. 1994, Luo 1994) have tested the first-year anisotropy maps from the *COBE* DMR experiment and find excellent statistical agreement with the hypothesis that the observed temperature fluctuations reflect random-phase Gaussian initial perturbations. However, since no competing models are examined, the compatibility with *non*-Gaussian models is not tested. In this Letter we employ the 2-point correlation function of extrema points to compare the 2-year DMR maps to a set of broadly applicable toy models employing both Gaussian and non-Gaussian statistics. Simulations employing inputs with known distributions indicate that this statistic can successfully distinguish Gaussian from non-Gaussian toy models with about 90% confidence, even at $10°$ angular resolution, and provide impetus for more computer-intensive studies of specific non-Gaussian cosmological models.

# 2. Analysis

We test for Gaussian statistics using the set of extrema points in the temperature field $T(\theta, \phi)$, defined as those points for which $\nabla T = 0$. For a pixelized map, this reduces to the collection of pixels hotter or colder than all of their nearest neighbors. Specifying pixels hotter than their neighbors produces a set of "hot spots" or "peaks", while specifying colder pixels produces "cold spots" or "valleys". An additional data selection may be performed, requiring $|T|$ to be greater than some threshold $\nu$, usually expressed in terms of the standard deviation $\sigma$ of the temperature field.

The 2-point correlation function of the extrema pixels provides a compact description of the data,

$$C_{ext}(\theta) = \frac{\sum_{i,j} w_i w_j T_i T_j}{\sum_{i,j} w_i w_j},$$

where $w$ is some weighting factor and the sum runs over all pixel pairs $\{i, j\}$ separated by angle $\theta$. We consider three applications of the extrema correlation function: peak-peak (autocorrelation of just the peaks or just the valleys), peak-valley (cross-correlation of the peak pixels with the valley pixels), and combined extrema (autocorrelation of all extrema points without regard for their second derivative). Bond & Efstathiou (1987) provide analytic approximations for these functions for random Gaussian fields but do not explicitly include the effects of instrument noise superposed on the CMB. Since the correlation properties of the non-uniform noise in the DMR maps are different from the underlying CMB temperature field, we use Monte Carlo techniques instead to derive the mean extrema correlation function and covariance as a function of the threshold $\nu$.

We analyze the extrema correlation functions of the 2-year *COBE* DMR maps (Bennett et al. 1994) and compare the sensitive 53 GHz (A+B)/2 sum maps and (A-B)/2 difference maps to Monte Carlo simulations of scale-invariant ($n = 1$) CMB anisotropy superposed with instrument noise. We generate each CMB realization using a spherical harmonic decomposition $T(\theta, \phi) = \sum_{\ell m} a_{\ell m} Y_{\ell m}(\theta, \phi)$ in which the harmonic coefficients $a_{\ell m}$ are random variables with zero mean and $\ell$-dependent variance

$$\langle a_{\ell m}^2 \rangle = (Q_{rms-PS})^2 \frac{4\pi}{5} \frac{\Gamma[l + (n-1)/2]\, \Gamma[(9-n)/2]}{\Gamma[l + (5-n)/2]\Gamma[(3+n)/2]}$$

(Bond & Efstathiou 1987). The coefficients $a_{\ell m}$ are drawn from parent populations with either Gaussian, log-normal, or $\chi_N^2$ ($N = 1$, 5, or 15 degrees of freedom) distributions, normalized to the mean and variances above. A non-Gaussian amplitude distribution for the $a_{\ell m}$ while retaining random phases provides a simple modification to the standard Gaussian model of CMB anisotropy. Although the non-Gaussian *amplitude* distributions tested here are skew-positive, the resulting sky maps are the convolution of the $a_{\ell m}$ with the spherical harmonics $Y_{\ell m}$ and are thus characterized by a negative kurtosis in the distribution of temperatures $T$ (e.g., higher "wings" than a Gaussian distribution). We test the sensitivity of our results to the transformation $a_{\ell m} \to -a_{\ell m}$ and find no difference using either definition.

The coefficients $a_{\ell m}$ define toy models to which specific models of structure formation may be compared (e.g. Weinberg & Cole 1992). Cosmological models with rare high-amplitude peaks, typified by topological defect models such as strings or texture, can be compared to the log-normal or $\chi_1^2$ distributions that tend to produce such features. The log-normal distribution is the most strongly non-Gaussian, while the $\chi_N^2$ models provide a smooth transition from strongly non-Gaussian ($N$=1) to nearly Gaussian ($N$=15). The models tested are not an exhaustive set of non-Gaussian models but are a computationally simple test of the power of various statistics on large angular scales.

We generate 1000 $n = 1$ full-sky realizations for each CMB model. To each CMB realization we add a realization of instrument noise defined by the level and pattern of noise in the DMR 2-year 53A and 53B channels (Bennett et al. 1994), then combine the channels to form (A+B)/2 sum maps and (A-B)/2 difference maps. We do not include Galactic emission or systematic uncertainties since these are small compared to the noise (Bennett et al. 1992, Kogut et al. 1992, Bennett et al. 1994).

We smooth the maps with a 7° Gaussian full width at half maximum (FWHM) as a compromise between suppressing noise and removing power at small scales, resulting in an effective smoothing on the sky of 10°. We reject pixels with Galactic latitude $|b|<20°$, remove fitted monopole and dipole temperatures from the surviving pixels, and determine the standard deviation $\sigma$. A nearest-neighbor algorithm then forms the collection of extrema pixels at thresholds $\nu = [0, 1, 2]\sigma$; these pixels at each threshold are then used to generate the peak-peak, peak-valley, and combined extrema correlation functions using unit weighting and $2°\!.6$ bins in the separation angle $\theta$. Since, by definition, two peaks can not be adjacent, we ignore the bin at zero separation and the first non-zero bin in all subsequent analysis. Analysis shows that the results are dominated by the first few remaining bins; consequently, we speed processing by truncating the correlation function at separation $\theta = 60°$.

The correlation functions at thresholds $\nu = [0, 1, 2]\sigma$ define a vector $D$, which we use to compare the DMR data to Monte Carlo simulations via a Gaussian approximation to the likelihood

$$\mathcal{L}(Q_{rms-PS}, \Upsilon) = (2\pi)^{-k/2} \frac{\exp(-\frac{1}{2}\chi^2)}{\sqrt{\det(\mathbf{M})}}$$

where $\chi^2 = \sum_{ij} (D_i - \langle D_i \rangle) (\mathbf{M}^{-1})_{ij} (D_j - \langle D_j \rangle)$, $\langle D \rangle$ is the simulation mean, and $\mathbf{M}$ is the covariance matrix between the $k$ bins of the vector $D$. The likelihood $\mathcal{L}$ is a function of two parameters: the continuous variable $Q_{rms-PS}$ representing the normalization and the discrete variable $\Upsilon$ representing the 5 $a_{\ell m}$ toy models. Since the covariance matrix $\mathbf{M}$ depends strongly on these parameters, a simple $\chi^2$ minimization approach fails. We evaluate the likelihood $\mathcal{L}$ in the 2-dimensional parameter space for values of $Q_{rms-PS}$ spanning the range $[0, 30]$ $\mu$K and search for the maximum in the resulting distribution.

## 3. Results and Discussion

Figure 1 shows the likelihood of the 2-year 53 GHz (A+B)/2 maps for both the Gaussian and the non-Gaussian models derived from the autocorrelation function of all extrema pixels. The likelihood function is greatest for the exact Gaussian model, with relative likelihoods for the non-Gaussian models $\sim 0.05$. Restricting the analysis to the $n = 1$ Gaussian models yields a maximum likelihood normalization for the 53 GHz (A+B)/2 maps of $Q_{rms-PS} = 18.1 \pm 1.9$ $\mu$K, in agreement with other estimates of $Q_{rms-PS}$ using the 2-year *COBE* data (Wright et al. 1994, Górski et al. 1994, Bennett et al. 1994, Banday et al. 1994). The width of a distribution in $Q_{rms-PS}$ is similar for all models. The likelihood function of the (A-B)/2 difference maps peaks at $Q_{rms-PS} = 0$ with no significant preference between models. Similar results occur for the peak-peak and peak-valley correlation functions.

The small likelihoods in Figure 1 for the non-Gaussian toy models given the DMR data would seem to rule out these models at high statistical confidence ($>99\%$).

However, formal identification of confidence intervals relies heavily on assumptions of the statistical distributions in the analysis (e.g. that the residuals $D - \langle D \rangle$ are multivariate normal) which are not always realized in practice. Furthermore, since the parameter $\Upsilon$ represents a collection of discrete models instead of a continuous variable, we can not integrate over $\Upsilon$ to derive confidence intervals in the usual way. We resolve these problems, assign formal confidence intervals, and test for statistical bias using a Monte Carlo approach.

We use the same machinery (likelihood analysis of the extrema correlation function) to generate the likelihood $\mathcal{L}(Q_{rms-PS}, \Upsilon)$ for 5000 simulated skies with $Q_{rms-PS}=18$ $\mu$K, 1000 realizations for each of the 5 toy models. For each realization we determine the maximum likelihood $\mathcal{L}(\Upsilon)$ evaluated at each model, and study the resulting distribution of likelihood maxima. Table 1 shows the percentage of simulations with maximum likelihood falling under each model. When the input is known to be Gaussian (column 2), 61% of the simulations correctly identify the exact Gaussian as the "best" model, with the remainder incorrectly allocated among the non-Gaussian models (a type I error). When the input is one of the non-Gaussian toy models instead (columns 3–6), that model is correctly identified in a similar fraction of the realizations with the *caveat* that the $\chi_1^2$ and log-normal distributions are nearly degenerate, so their contributions should be added in each column. There is no evidence for any statistical bias favoring one particular model. We may thus quantify the confidence intervals in terms of type II error (accepting a hypothesis when it is false): given that the DMR likelihood is greatest for the Gaussian model, how confident are we that the CMB is *not* in fact a realization of one of the non-Gaussian toy models? From the first row of Table 1 (Gaussian model received highest likelihood) we see that the probability of obtaining this result is three times larger for a Gaussian CMB than for any of the non-Gaussian toy models. We may thus reject the non-Gaussian toy models at 75% confidence.

A more powerful test uses additional information from the likelihood distribution. We have examined the sub-set of simulations for which the best fitted model was not, in fact, the correct input, and found that the likelihoods in these cases did not strongly select against the rejected models. The DMR likelihood does not show this pattern: the second-best likelihood (for the $\chi_{15}^2$ model) is only 0.08. Table 2 shows the percentage of simulations meeting 2 conditions: that the likelihood peaks in a selected output model (as in Table 1), and that the next-best likelihood be smaller than 0.08. We recover the same overall pattern as Table 1: the most probable outcome is to recover the input model correctly, but the fraction of both type I errors (columns) and type II errors (rows) is reduced. The probability of obtaining a result similar to the DMR likelihood is approximately ten times greater for the Gaussian CMB model than the non-Gaussian toy models, allowing us to reject these models at 90% confidence.

The topological quantity known as the genus has also been proposed as a test for non-Gaussian statistics in the CMB (Gott et al. 1990). Smoot et al. (1994) show that the genus of the first-year DMR maps is consistent with random-phase Gaussian models. We have performed a likelihood analysis of the genus of the 2-year DMR maps compared to the same set of non-Gaussian toy models used for the extrema correlation analysis above. Although the genus likelihood is also greatest for the exact Gaussian model, the ability to reject the non-Gaussian models is weaker, with maximum likelihood $\text{Max}(\mathcal{L}) \approx 0.3$ for the log-normal and $\chi_1^2$ models using the genus compared to $\text{Max}(\mathcal{L}) \approx 0.05$ using the extrema correlation function. The genus of the 2-year DMR maps will be discussed in greater detail in a future paper.

Both the genus and the extrema correlation function show the 2-year DMR data to be consistent with the hypothesis of random-phase Gaussian statistics, and inconsistent at the 90% confidence level with random-phase toy models with non-Gaussian distributions of the spherical harmonic coefficients $a_{\ell m}$. The ability to reject non-Gaussian models is limited by type II errors and reflects the generally larger role of cosmic variance in the non-Gaussian toy models tested in this paper. Although the statistical power of these tests is not overwhelming, it does demonstrate that large-beam experiments can probe the statistical distribution of CMB anisotropy. Physically motivated non-Gaussian models (e.g. topological defects) have strong phase correlations as well, which would be expected to increase the statistical power of these tests. There is thus an incentive to pursue further tests of specific models using the *COBE* DMR data.

We gratefully acknowledge the dedicated efforts of those responsible for the *COBE* DMR data. C. Lineweaver and L. Tenorio provided helpful discussion of statistical techniques. *COBE* is supported by the Office of Space Sciences of NASA Headquarters.

Table 1: Percentage of Simulations From Single Test

| Fitted Model | Input Model[a] | | | | |
|---|---|---|---|---|---|
| | Gaussian | $\chi^2_{15}$ | $\chi^2_5$ | $\chi^2_1$ | Log-Normal |
| Gaussian | 61 | 21 | 20 | 23 | 24 |
| $\chi^2_{15}$ | 16 | 56 | 21 | 21 | 23 |
| $\chi^2_5$ | 11 | 11 | 44 | 15 | 18 |
| $\chi^2_1$ | 5 | 6 | 7 | 28 | 15 |
| Log-Normal | 7 | 6 | 8 | 13 | 20 |

[a] Percentage of 1000 simulations of each input model whose likelihood maxima falls under each output model.

Table 2: Percentage of Simulations From Double Test

| Fitted Model | Input Model[a] | | | | |
|---|---|---|---|---|---|
| | Gaussian | $\chi^2_{15}$ | $\chi^2_5$ | $\chi^2_1$ | Log-Normal |
| Gaussian | 20 | 2 | 2 | 3 | 4 |
| $\chi^2_{15}$ | 1 | 16 | 1 | 3 | 3 |
| $\chi^2_5$ | 1 | 1 | 10 | 1 | 2 |
| $\chi^2_1$ | 2 | 2 | 3 | 15 | 8 |
| Log-Normal | 4 | 4 | 5 | 7 | 13 |

[a] Percentage of 1000 simulations of each input model whose likelihood maxima falls under each output model, and whose next-best likelihood is less than 0.08.

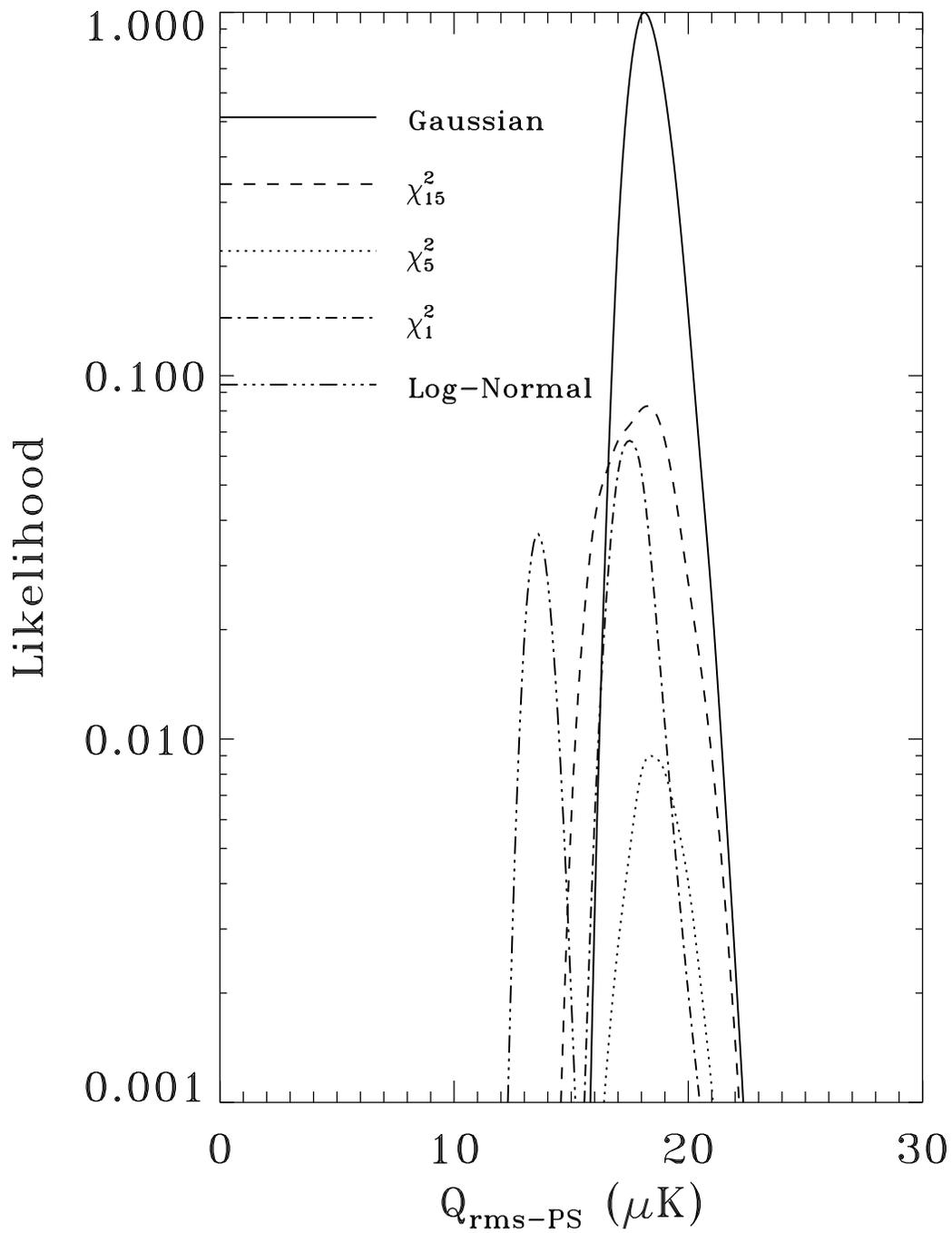

Figure 1: Likelihood function of the 2-year DMR 53 GHz (A+B)/2 extrema correlation function for Gaussian and non-Gaussian toy models.